\documentstyle[fleqn,11pt]{article}
\date{}

\begin{document}
\title{\Large \bf The global colour model of QCD and its 
      relationship to the NJL model, chiral perturbation theory and other models
\thanks{Contribution to the {\it Joint Japan Australia Workshop} on {\bf Quarks,
Hadrons and Nuclei}, Adelaide, November 1995.}}  
\author{{Reginald T. Cahill  and      Susan M. Gunner}\\
  {Department of Physics, Flinders University
\thanks{E-mail: Reg.Cahill@flinders.edu.au,
Susan.Gunner@flinders.edu.au}}\\ { GPO Box 2100, Adelaide 5001, Australia
}\\ {and Institute of Theoretical Physics, Adelaide}\\{hep-ph/9601319}}

\maketitle

\begin{center}
\begin{minipage}{120mm}
\vskip 0.6in
\begin{center}{\bf Abstract}\end{center}
{The Global Colour Model (GCM) of QCD is a very successful model. Not only is it
formally derivable from QCD but under various conditions it reduces to the NJL
model and also to Chiral Perturbation Theory, and to other models. Results 
presented  include the effective gluon propagator,  the difference between
constituent   and  exact quark propagators, various meson and
nucleon observables,   a new form for the mass formula for the Nambu-Goldstone
mesons of QCD, and the change in the MIT bag constant in nuclei. }

\end{minipage} \end{center}

\noindent {\bf 1. The  Global Colour Model  (GCM)   }  

The GCM can be formally 
derived from QCD \cite{RTC}. The GCM is in turn  easily related to a number of
the  more phenomenological models of QCD as indicated in  fig.1.

\begin{center}
\begin{picture}(10,150)(+200,80)
\thicklines
\put(5,185){\large \bf QCD}
\put(25,165){\vector(0,-1){30}}
\put(55,190){\vector(3,0){30}}
\put(95,185){\large \bf GCM}
\put(155,190){\vector(3,4){30}}
\put(200,225){\large \bf NJL}
\put(155,190){\vector(3,0){30}}
\put(200,185){\large \bf ChPT}
\put(155,190){\vector(3,-4){30}}
\put(200,155){\large \bf MIT, Cloudy Bag,  }
\put(200,140){\large \bf Soliton Models, QHD, QMC....}
\put(29,115){\large \bf Lattice Gluons}
\put(150,120){\vector(1,0){40}}
\put(120,135){\vector(0,1){30}}
\put(200,115){\large \bf Lattice Hadrons}
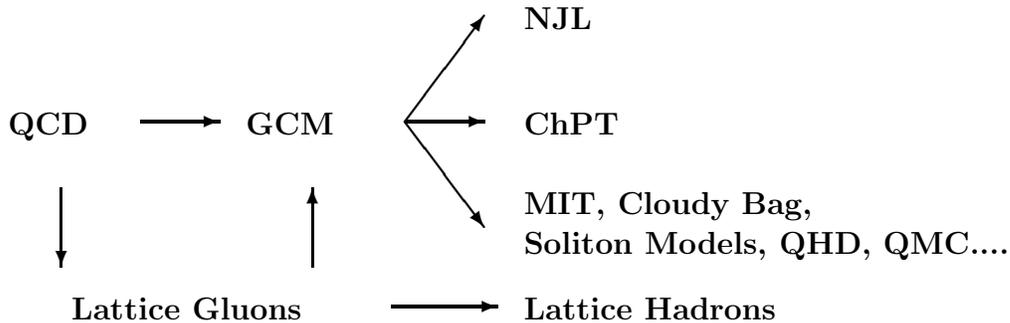
\put(10,90){  Figure1  Relationship of the GCM to QCD and other models}  
\end{picture}
\end{center}
QCD and models for  QCD amount to the calculation of correlation functions.

\begin{equation}
{\cal G}(..,x,...)=\int{\cal D}\overline{q}{\cal D}q{\cal
DA}....q(x).....exp(-S_{qcd}[A,\overline{q},q])
\end{equation}
At low energies we expect that only hadronic correlations are directly
observable, and  we expect  (1) to be transformable into an hadronic
form, equivalent to the effective action description of low energy nuclear
physics. 

\begin{eqnarray*}
\int {\cal D}\overline{q}{\cal D}q{\cal
D}Aexp(-S_{qcd}[A,\overline{q},q]+\overline{\eta}q+ \overline{q}\eta)
\approx \end{eqnarray*}
\begin{equation}  \mbox{\ \ \ \ \ \ \ \ \ \ \ \ \ \ \ \ \ \ \ \ } 
\int{\cal D}\pi{\cal D}\overline{N}{\cal
D}N...exp(-S_{had}[\pi,...,\overline{N},N,..]+J_{\pi}[\overline{\eta},\eta]\pi+..)
\end{equation}
If in (1) we formally do the gluon integrations, after fixing a gauge (ghosts not
shown)

\begin{equation}
\int {\cal D}\overline{q}{\cal D}q{\cal
D}Aexp(-S_{qcd}[A,\overline{q},q]+\overline{\eta}q+ \overline{q}\eta)=
\int {\cal D}\overline{q}{\cal D}q
exp(-S[\overline{q},q]+\overline{\eta}q+ \overline{q}\eta),\end{equation}
leaving an action for quarks of the form
\begin{equation}
S[\overline{q},q]= \int 
\overline{q}(\gamma . \partial+{\cal M}) q+\frac{1}{2}\int
j^a_{\mu}(x)j^a_{\nu}(y)D_{\mu\nu}(x-y)+\frac{1}{3!}\int
j^a_{\mu}j^b_{\nu}j^c_{\rho}D^{abc}_{\mu\nu\rho}+...
\end{equation}
where $j^a_{\mu}(x)=\overline{q}(x)\frac{\lambda^a}{2}\gamma_{\mu}q(x)$, and 
$D_{\mu\nu}(x)$ is the exact pure gluon propagator 
\begin{equation}D_{\mu\nu}(x)=\int {\cal
D}AA^a_{\mu}A^a_{\nu}exp(-S_{qcd}[A,0,0]) \end{equation}
A variety of techniques for computing $D_{\mu\nu}(x)$ exist, such as
Dyson-Schwinger  equations (DSE)
\cite{Pennington}, lattice simulations \cite{Marenzoni93,Marenzoni94} and, as
discussed in other papers at the workshop, monopole and instanton modellings.  
Dropping the
$D^{abc}_{\mu\nu\rho},..$ defines the GCM.   This is equivalent to using a
quark-gluon field theory with the action.
\begin{equation}
S_{gcm}[\overline{q},q,A^a_{\mu}]=\int \left( 
\overline{q}(\gamma . \partial+{\cal
M}+iA^a_{\mu}\frac{\lambda^a}{2}\gamma_{\mu})q
 +\frac{1}{2}
A^a_{\mu}D^{-1}_{\mu\nu}(i\partial)A^a_{\nu} \right)
\end{equation}
where $D^{-1}_{\mu\nu}(p)$ is the matrix inverse of $D_{\mu\nu}(p)$.
This action has  a global colour symmetry, hence the description as the GCM.
Hadronisation \cite{RTC} of the GCM involves a sequence of functional integral 
calculus changes of variables involving, in part, the transformation to bilocal
meson and diquark fields, and then to the usual local meson and baryon fields.

\begin{equation}\int {\cal D}\overline{q}{\cal D}q{\cal
D}Aexp(-S_{qcd}[A,\overline{q},q])\end{equation}
\begin{equation} \mbox{\ \ \ \ \ \ \ \ \ \ \ \ }\approx \int {\cal D}\overline{q}{\cal D}q
exp(-\int 
\overline{q}(\gamma . \partial+{\cal M})q +\frac{1}{2}\int
j^a_{\mu}(x)j^a_{\nu}(y)D_{\mu\nu}(x-y))\end{equation}

\begin{equation} \mbox{\ \ \ \ \ \ \ \ \ \ \ \ }
=\int D{\cal B}D{\cal D}D{\cal D}^{\star}exp(-S[{\cal B}, {\cal D}, {\cal
D}^{\star}]) \mbox{\ \ \ \ (bilocal fields)} \end{equation} 
\begin{equation} \mbox{\ \ \ \ \ \ \ \ \ \ \ \ } =\int{\cal D}\pi{\cal D}\overline{N}{\cal
   D}N...exp(-S_{had}[\pi,...,\overline{N},N,..]) 
  \mbox{\ \ \ \ (local fields) }\end{equation}
The derived hadronic  action, to low order in fields and derivatives, has the form
\begin{eqnarray*} \lefteqn{ S_{had}[\pi,...,\overline{N},N,..] = }\\&
 &\mbox{\ \ \ \ \ \ \ \ \ \ \ \ \ \ }\int d^4x
tr\{\overline{N}(\gamma.\partial+m_N+  \Delta m_N- 
m_N\surd 2i\gamma_5\pi^a{\cal T}^a+..)N\}+\\
& &\mbox{\ \ \ \ \ \ \ \ \ \ }+\int d^4x\left[ \frac{f_{\pi}^2}{2}
[(\partial_{\mu}\pi)^2+m_{\pi}^2\pi^2] + \frac{f_{\rho}^2}{2}[ -
\rho_{\mu}\Box\rho_{\mu} +(\partial_{\mu}\rho_{\mu})^2 +m_{\rho}^2\rho_{\mu}^2]+\right.\\   
& &\left.\mbox{\ \ \ \ \ \ \ \ \ \ }+\frac{f_{\omega}^2}{2}[\rho
\rightarrow \omega]-f_{\rho}f_{\pi}^2g_ {\rho\pi\pi}\rho_{\mu}.\pi \times\partial
_{\mu}\pi
-if_{\omega}f_{\pi}^3\epsilon_{\mu\nu\sigma\tau}\omega_{\mu}\partial_{\nu}
 \pi . \partial_{\sigma}\pi\times \partial_{\tau}\pi+\right.\\
& &\left.\mbox{\ \ \ \ \ \ \ \ \ \ }-if_{\omega}f_{\rho}f_{\pi}
G_{\omega\rho\pi}\epsilon_{\mu\nu\sigma\tau}\omega_{\mu}\partial_{\nu}
\rho_{\sigma}.\partial_{\tau}\pi+\right.\end{eqnarray*}
\begin{equation}\left.\mbox{\ \ \ \ \ \ \ \ \ \ \ \ \ \ \ \ }+
\frac{\lambda
i}{80\pi^2}\epsilon_{\mu\nu\sigma\tau}tr(\pi.F\partial_{\mu}
\pi.F\partial_{\nu}\pi.F\partial_{\sigma}\pi.F\partial_{\tau}\pi. F)
+......\right]\end{equation}
This induced effective action is the action of Quantum Hadro-Dynamics (QHD).
Being a derivative expansion it should not be used in hadronic loop
calculations. For that purpose the non-local form of the hadronic effective
action is necessary.  The above derivation leads naturally
to the dynamical breaking of chiral symmetry (sect.2) and so to the Chiral
Perturbation Theory phenomenology (ChPT) (sect.4). The bare mesons are described
by Bethe-Salpeter equations (BSE), and the baryons by covariant Faddeev equations
(FE) \cite{RTC}. The GCM thus relates hadronic properties directly to the gluon
$D_{\mu\nu}$. It is not known why the GCM truncation is so effective.

\vspace{5mm}

\noindent   {\bf 2. Dynamical Chiral Symmetry Breaking} 

The hadronic effective
action in (10) arises from expanding the bilocal action in (9) about its minimum
via the Euler-Lagrange equations (ELE):
$\delta S/\delta {\cal B}=0$ give the  rainbow constituent quark DSE in (12)
and (13), while
$\delta S/\delta {\cal D}=0$ has solution ${\cal D}=0$. Fluctuations are
described by the  curvatures: 
$\delta^2 S/\delta {\cal B}\delta {\cal B}$ gives the ladder BSE for mesons, and
$\delta^2 S/\delta {\cal D}\delta {\cal D^*}$ gives the ladder BSE for diquarks.

\begin{equation}B(p)=\frac{16}{3}\int\frac{d^4q}{(2\pi)^4}D(p-q)
\frac{B(q)+m}{q^2A(q)^2+(B(q)+m)^2}\end{equation}

\begin{equation}[A(p)-1]p^2=\frac{8}{3}\int\frac{d^4q}{(2\pi)^4}q.pD(p-q)
\frac{A(q)}{q^2A(q)^2+
(B(q)+m)^2}\end{equation}

\begin{equation}G(q)=(iA(q)q.\gamma+B(q)+m)^{-1}=-iq.
\gamma\sigma_v(q)+\sigma_s(q)\end{equation}
Note that the quark propagator in (14) is a constituent quark propagator (and
appears in the exponent in (9)),  and not the exact quark propagator, as would
arise from (1), and  which  would satisfy the Slavnov-Taylor identities (STI). In
the GCM and related phenomenologies the quarks in the bare hadrons in (11) are
further dressed, for example by mesons, in the functional integrations of (10).
Imposing the STI in (12) and (13) could thus result in double counting
problems. The exact quark propagator is relevant to the question of absolute quark
confinement.  The constituent quark propagator has the running mass
$M(q)=B(q)/A(q)$. The constituent quark mass arises as the most likely value of
$M(q)$ in, say, the BSE. It will depend on the bare current mass and to some extent even the hadron state.

\vspace{5mm}

\noindent   {\bf 3. The Nambu - Jona-Lasinio  Model (NJL)} 

 The NJL model is a
special case of the GCM.  Formally the NJL model \cite{Reinhardt90} is the contact
interaction limit of the GCM:
$ D_{\mu\nu}(x-y)\rightarrow g^2\delta_{\mu\nu}\delta(x-y)$  or $ D(p) \rightarrow
g^2$. But in the constituent quark  DSE  the contact limit  is undefined because it
leads to divergences in (12) and (13).  A cutoff $\Lambda$ is then always
introduced in NJL computations, which is equivalent to  using the gluon
propagator $ D(p)
\rightarrow g^2\theta(q^2-\Lambda^2) $.  Hence the NJL model is the  GCM  but with
a box-shaped  $D(p)$, rather than a `running'  $D(p)$.  The GCM bosonisation
in (8) arises from a generalised Fierz identity \cite{RTC}
\begin{equation}
j^a_{\mu}(x)j^a_{\mu}(x)=
 \overline{q}(x)\frac{M^{\theta}_m}{2}q(x)\overline{q}(x)\frac{M^{\theta}_m}{2}q(x)
 -\overline{q}(x)\frac{M^{\phi}_d}{2}\overline{q}(x)
  ^{cT}q(x)^{cT}\frac{M^{\phi}_d}{2}q(x)\end{equation}
where $\{M^\theta_m\}=\{\surd\frac{4}{3}K^a\otimes 1 \otimes F^c\}$
and $\{M^{\phi}_d\}=\{i\surd\frac{2}{3}K^a\otimes\epsilon^{\rho}\otimes H^f\}$
involve generators of the Dirac, colour and flavour symmetries.  The use of this
identity gives the generalized NJL model.

\vspace{5mm}

\noindent  {\bf 4. Chiral Perturbation Theory (ChPT)}

 When the  quark current
masses ${\cal M} \rightarrow 0$   $S[\overline{q},q,A^a_{\mu}]$ has  an
additional global $U_L(N_F)\otimes U_R(N_F)$ chiral symmetry:  writing 
$\overline{q}\gamma_{\mu}q=\overline{q}_R\gamma_{\mu}q_R+
\overline{q}_L\gamma_{\mu}q_L$ 
where $q_{R,L}=P_{R,L}q$ and $\overline{q}_{R,L}=\overline{q}P_{L,R}$ we see that 
these two   parts are separately invariant under 
$q_R\rightarrow U_Rq_R,
\overline{q}_R\rightarrow\overline{q}_RU^{\dagger}_R$ and $q_L\rightarrow U_Lq_L,
\overline{q}_L\rightarrow\overline{q}_LU^{\dagger}_L$. 
Its consequences may be explicitly traced through the GCM hadronisation.
First the ELE $\delta S/\delta {\cal B}$=0 have degenerate
solutions. In terms of the constituent quark propagator we find
 \begin{equation}G(q;V)=[iA(q)q.\gamma+VB(q)]^{-1}=\zeta^\dagger G(q;{\bf
1})\zeta^\dagger\end{equation}
where 
 \begin{equation}\zeta=\surd V, \mbox{\ \  } V=exp(i\surd2\gamma_5\pi^aF^a)\end{equation}
and consequently the fluctuations $\delta^2 S/\delta {\cal B}\delta {\cal B}$ have
massless  Nambu-Goldstone (NG) BSE states.

 In the hadronisation (10) new variables are forced upon us to describe the
degenerate minima (vacuum manifold): 
 \begin{equation}U(x)=exp(i\surd 2\pi^a(x)F^a)\end{equation} 
\begin{equation}V(x)=P_LU(x)^{\dagger}+P_RU(x) =exp(i\surd 2\gamma_5\pi^a(x)F^a)\end{equation}

  The NG part of the hadronisation gives
\[
 \int d^4x\left( \frac{f_{\pi}^2}{4}tr(\partial_{\mu}U\partial_{\mu}U^
{\dagger})+\kappa_1tr(\partial^2U\partial^2U^{\dagger})+
\frac{\rho}{2}tr([{\bf 1} -\frac{U+U^{\dagger}}{2}]{\cal M})+\right.\]   
\begin{equation}\mbox{\ \ \ \ \ \ \ \ \ \ \ \ \ \ \ \ \ \ \ \ \ \ \ \ \ \ \ \ \ \ }
\left. +\kappa_2tr([\partial_{\mu}U\partial_{\mu}U^{\dagger}]^2)
+\kappa_3tr(\partial_{\mu}U\partial_{\nu}U^{\dagger}\partial_{\mu}U \partial_{\nu}U^
{\dagger})+....\right) \end{equation}
This is the ChPT  effective action \cite{CPT}, but with the added insight that all
coefficients are given by explicit and convergent integrals in terms of $A$ and
$B$, which are in turn  determined by $D_{\mu\nu}$.  The higher order terms
contribute to
$\pi\pi$ scattering.  The dependence of the ChPT coefficients upon  $D_{\mu\nu}$
have been studied  in \cite{Pi94,CG95,Frank95}.
The hadronisation procedure gives a full account of NG-meson - nucleon
coupling.

\vspace{5mm}

\noindent  {\bf 5. The MIT, Cloudy Bag and  Soliton Models} 

 While the GCM
hadronisation in (11) and (20) is the  main result, at an intermediate stage one
obtains \cite{CR85} extended meson  Quark-Meson Coupling type models (QMC)
\cite{G88}. Applying mean field techniques to this GCM quark-meson coupling 
effective action leads to soliton type models, which have been studied in detail in
\cite{Frank90} and the significance of the extended mesons demonstrated. From the
soliton  models a futher ansatz for the soliton form leads to  the MIT and Cloudy
Bag Model (CBM).  In particular the GCM expression for the MIT Bag constant may be
extracted:

\begin{equation}{\cal B}= \frac{12\pi^2}{(2\pi)^4}\int_0^\infty
sds[\mbox{ln}(\frac{A^2(s)s+B^2(s)}{A^2(s)s})
-\frac{B^2(s)}{A^2(s)s+B^2(s)}]\end{equation} which is based on the energy density
for complete restoration of chiral symmetry inside a cavity. This bag constant is
for   core states as no meson cloud effect is included.

With a mean field description of the pion sector  via $\sigma(x)$, 
which describes the isoscalar part of
$\sigma(x)V(x)$, where $\sigma(x)$ is a `radial'
field multiplying the NG boson field $V(x)$  

\begin{equation}{\cal B}(\sigma)= \frac{12\pi^2}{(2\pi)^4}\int_0^\infty
sds[\mbox{ln}(\frac{A^2(s)s+\sigma^2B^2(s)}{A^2(s)s})
-\frac{\sigma^2B^2(s)}{A^2(s)s+B^2(s)}]\end{equation}
which reduces to (21) when 
$\sigma=1$, being the non-perturbative field external to an isolated  nucleon
core, and $\sigma<1$ describing a partial restoration of chiral symmetry outside of
the core. Using the gluon propagator discussed in sect.6 we obtain the plot of
${\cal B}(\sigma)/{\cal B}(0)$ shown in fig.2. Dressing of the nucleon core by
mesons is partly described by  a reduction in $\sigma$ in the surface region,
causing a reduction in the nucleon mass.

 However in nuclei a mean meson field   description \cite{ST} means that
$\sigma$  is even further reduced  outside of the nucleons, and the effective bag
constant is further reduced. The $\sigma$ field, which can  model in part
correlated $\pi\pi$ exchanges,  along with the $\omega$ meson field, are believed
to be important to a mean field modelling of nuclei.  In \cite{JJ} it has been
argued that the  reduction of the effective bag constant  for nucleons inside
nuclei  is essential  to the recovery of features of relativistic nuclear
phenomenology.  The GCM thus allows ${\cal B}(\sigma)$ and details of relativistic
nuclear phenomenology to be directly related to the constituent quark propagator,
and in turn to the gluon propagator.

\vspace{5mm}

\noindent  {\bf 6. Gluon Propagator  Separable Expansions} 

 In application the 
GCM  model amounts to solving a sequence of non-linear and linear integral
equations: 

$D_{\mu\nu} \mbox{\ } \rightarrow \mbox{\ }G_{const.quark}\mbox{\ } 
\rightarrow meson \mbox{\ } BSE \mbox{\ } \& \mbox{\ }diquark \mbox{\ } BSE
\rightarrow$
 
$\mbox{\ }covariant \mbox{\ } Faddeev
\mbox{\ } eqns\mbox{\ } for\mbox{\ } baryons \rightarrow \mbox{\ } hadronic
\mbox{\ } observables$ 

In GCM computations one uses what amounts to a mixed metric:  a Euclidean metric
for the  internal quark-gluon fluctuations, and with the hadron momenta in the
time-like region of the Minkowski metric (usually but not necessarily in the
centre-of-mass (cm) frame). The constituent quark propagator equations (12) and
(13) are solved in the Euclidean region ($s=q^2 \geq 0$), and then  $A(q^2)$ and
$B(q^2)$ are $O(4)$ invariant. Significantly it was discovered \cite{SC94} that
when low mass BSE states are solved with this mixed metric approach  the
(on-mass-shell) bound state form factors also show a remarkable degree of $O(4)$
invariance wrt the relative quark momentum, even though this was not assumed in
the numerical solution of the BSE.  If we note that an $O(4)$ hyperspherical
expansion \begin{equation}
D(p-q)=D_0(p^2,q^2)+q.pD_1(p^2,q^2)+...
\end{equation}
of the gluon propagator, where 
\begin{equation}
D_0(p^2,q^2)=\frac{2}{\pi}\int_0^{\pi}d\beta\,sin^2\beta\, 
D(p^2+q^2-2 p q cos\beta)
\end{equation}
together with the  numerical technique of using a multi-rank separable expansion
\begin{equation}
D_0(p^2,q^2)=\sum_{i=1,n} \Gamma_i(p^2)\Gamma_i(q^2),......
\end{equation}
then this separable expansion automatically generates $O(4)$ invariant  BSE
solutions when (23) is used, at lowest order, in a BSE.  Hence the significant
realisation that using a  multi-rank separable expansion is ideally suited for
GCM computations.  This technique then renders the DSE and BSE equations to
essentially algebraic form, with only the baryon integral equation computations
requiring extensive numerical solution. As shown in \cite{SC94} this very
useful property of $O(4)$ BSE solution invariance appears to be a consequence of
having no nearby singularities in the constituent quark propagator, and so appears
to be related to the confinement property of the quarks.  

It is important to note that the separable technique is introduced  only as a
numerical technique and only after all integral equations have been formally
derived using the covariance of the GCM. If the separable expansion of the gluon
propagator were  introduced  in the defining action  of the GCM then the
explicit breaking  of covariance would  block any of the usual momentum-space 
DSE, BSE,.. equations from being derivable.

 Implementing  the separable expansion \cite{CG95} we first solve the DSE (12) and
(13). Then sums are obtained
\begin{equation}
B(s)=\sum b_i\Gamma_i(s), ...\mbox{\ \ \ \ }
\sigma_s(s)=\sum_{i=1,n}\sigma_s(s)_i, \mbox{ \ \ \ \ }
\sigma_v(s)=\sum_{i=1,k}\sigma_v(s)_i  
\end{equation} 
 The gluon propagator is implicitly parametrised in \cite{CG95} by assuming entire
functions (as a means of implementing quark confinement) for $\sigma_s$
and $\sigma_v$, say
\begin{equation}
\sigma_s(s)_i=c_iexp(-d_is), \mbox{ \ \ \ \ }
\sigma_v(s)=  \frac{2s-\beta^2(1-exp(-2s/\beta^2))}{2s^2}  
\end{equation}
Then  (12) gives
\begin{equation}
 b_i^2=\frac{16}{3}\pi^2\int_0^{\infty} sds B(s)_i\frac{B(s)}{sA(s)^2+B(s)^2}
\end{equation}
in which $B(s)=B(s)_1+B(s)_2+..$, and
$B(s)_i=\sigma_s(s)_i/(s\sigma_v(s)^2+\sigma_s(s)^2)$.

Having solved the constituent quark non-linear equations (12) and (13) one can
then proceed to solve the meson and diquark BSE and finally the constituent
nucleon-core Faddeev equations \cite{RTC}.  The implicit gluon propagator 
parameters in (27) are then determined by a best fit   to some hadron data:
$f_{\pi},m_{\pi},m_{a_1}$. Subtleties associated with the incorporation of the
quark current masses and the reasons for choosing to fit these particular data are
discussed in \cite{CG95}. The gluon propagator parameter values are also given in 
\cite{CG95}, and various hadron observables are presented, and discussed here in
sect. 7. 

Having determined the best fit the translation invariant form of 
the gluon propagator may be determined  from (24) which, with $q^2=0$, gives 

\begin{eqnarray*} D(p^2)=D_0(p^2,0) =\sum_i\Gamma_i(p^2)\Gamma_i(0)
=\end{eqnarray*}
\begin{equation}\mbox{\ \ \ \ \ \ \ \ \ \ \ \ \ \ \ \ \ \ \ \ \ \ \ \ }
=\sum_i\frac{1}{b_i^2}B(p^2)_iB(0)_i=\sum_i\frac{1}{b_i^2}\frac{\sigma_s(0)_i}{\sigma_s(0)^2}\frac{\sigma_s(p^2)_i}
 {p^2\sigma_v(p^2)^2+\sigma_s(p^2)^2}\end{equation}
However the form of the extracted gluon propagator will clearly depend on the
forms assumed in  (27), and futhermore, while a unique best fit was obtained the
fit is somewhat flat wrt variation of some of the parameters.  To help
more accurately resolve  the form of the effective gluon propagator that arises
in the GCM, current work  \cite{CGS96} that constitutes a Hybrid 
Lattice-GCM calculation.  In this a multi-rank separable expansion is constructed
and its parameters determined by fits to both some hadron data and as well to the
form for the gluon propagator that arises from lattice computations
\cite{Marenzoni93,Marenzoni94}, except for the infrared (IR) region of the gluon
propagator.  This amounts to probing the deep IR properties of the gluon
propagator by means of low energy hadronic data.  Results will be reported
elsewhere.

\vspace{5mm}
\noindent   {\bf 7. Meson and Nucleon Observables} 

 Table 1 shows various
hadronic observables computed using the separable expansion technique  of sect. 6.
Of particular relevance here are the $\pi\pi$ scattering lengths which arise from
the ChPT effective action (20) with the parameters given by the explicit GCM 
forms \cite{Pi94}.  The constituent quark masses arise from the  value of the
constituent quark running mass $M(s)$ at the value corresponding to the dominant
$s$ value in the BSE. This can only be determined after the BSE solution is known. 
Various diquark masses are shown. Diquarks are extended quark-quark correlations
in  baryons, and arise naturally  from the GCM hadronisation \cite{RTC}. They are
particularly effective in reducing the numerical complexity when solving the baryon
Faddeev integral equations.  Note that a nucleon-core mass  of $1390$MeV
is consistent with the first Faddeev computation of the  nucleon-core mass in
\cite{Burden}. This is also consistent with the idea that dressing of the
nucleon-core  by NG bosons causes a decrease of some
$300$MeV.  Incorporation of the spin $1^+$ diquark state into the  nucleon-core
computation is also expected to decrease the nucleon mass.  Other computations of
the  nucleon-core state \cite{Buck,Ishii,Huang,Meyer} in the context of the NJL
limit of the GCM have always adjusted the NJL parameters so that the  nucleon-core
mass actually fitted the experimental nucleon mass, leaving out the significant NG
dressing effect.

Also shown is the GCM predicted value for the MIT bag constant,  using  expression
(21). This is somewhat larger then the usual MIT value and gives a MIT
nucleon-core mass of
$1500$MeV without  cm corrections. With cm corrections we might
expect this to reduce to near the covariant Faddeev value of $1390$MeV, i.e the
GCM MIT bag constant value appears to be consistent with the expected 
nucleon-core mass. The further reduction in the GCM ${\cal B}$ value when chiral
symmetry is partially restored due to meson dressing of isolated nucleons, and a
further enhanced reduction for nucleons inside nuclei, was discussed in sect.5.

\vspace{5mm} 

\noindent  {\bf 8. A  New NG Mass Formula} 

 The usual NG mass formula is
\begin{equation}
M_{\pi}^2= \frac{(m_u+m_d)\rho}{f_{\pi}^2}
\end{equation}
where
\begin{equation}
\rho=<\overline{q}q>=N_ctr(G(x=0))=12\int\frac{d^4q}{(2\pi)^4}\sigma_s(q^2)
\end{equation}
This expression for $\rho$  is
divergent in QCD.   The  values of $m$ and
$<\overline{q}q>$ are then usually quoted as being relative to  
some cutoff momentum, often $1$GeV.

 However analysis of the GCM \cite{CG96} gives rise to a different mass formula
in which 
\begin{equation}
\rho=24\int\frac{d^4q}{(2\pi)^4}\epsilon_s(q^2)c(q^2)\sigma_s(q^2) 
\end{equation}
with the   factor $c(s)=B(s)^2/(sA(s)^2+B(s)^2)$, and    the expansion  of
the $m$ dependence of $B(q)$ from (12)  \begin{equation}
B(q)\mid_{m\neq0}+m=B(q)\mid_{m=0}+\epsilon_s(q)m+O(m^2)
\end{equation}
defines the  current mass enhancement factor  $\epsilon_s(q)$.
Despite the apparent difference between (31) and (32) Langfeld and Kettner
\cite{Langfeld} have shown that they are indeed equal.

\vspace{5mm}

\noindent  {\bf 9. Conclusions} 

 We have summarised here some of the 
facets of the GCM and its relationship to the fundamental theory (QCD) and to
various phenomenological models which have been profitably employed in 
modelling QCD. Significantly the GCM permits not only the derivation of these
phenomenologies, but also particular expressions for the numerous parameters that
arise in these models, such as the MIT bag constant, its modification within
nuclear matter, and the ChPT parameters. Fundamentally the GCM is non-local and
so  the predictions do not involve divergences and the consequent
renormalisations, once the gluon propagator is given  its scale dependence.

The authors thank Professors K. Saito and  A.W. Thomas and Dr A.G. Williams 
for the organisation of this workshop.

\newpage

\begin{figure} [t]
\vspace{70mm}
\hspace{2mm}\special{illustration Bagvar.ps}
\end{figure}
Figure 2 Variation of MIT Bag constant ${\cal B}(\sigma)/{\cal B}(1)$  wrt
$\sigma$. Partial restoration of chiral symmetry   is described by $\sigma < 1$.

\vspace{5mm}

\hspace{5mm} \begin{tabular}{l r r}
\multicolumn{3} {l}  {\bf Table 1:  Hadronic Observables}\\
\hline \hline
 {\bf Observable} & {\bf Theory} & {\bf Expt./Theory} \\ \hline \hline  {\bf
Fitted observables}\\

$f_{\pi}$  &    93.00MeV  &93.00MeV\\
$a_1$ meson mass  & 1230MeV & 1230MeV\\
$\pi$ meson mass   &  138.5MeV   & 138.5MeV\\
$K $ meson mass (for $m_s$ only)   &  496MeV      &     496MeV\\
\hline {\bf Predicted observables}\\
$(m_u+m_d)/2$   &  6.5MeV  &  6.0MeV\\
$m_s$   &  135MeV   &  130MeV\\\
$\omega$ meson mass &  804MeV  &  782MeV  \\
$a^0_0$ $\pi-\pi$ scattering length &  0.1634  &  0.21 $\pm$ 0.01 \\
$a^2_0$ $\pi-\pi$ scattering length&  -0.0466  &  -0.040 $\pm$ 0.003 \\
$a^1_1$  $\pi-\pi$ scattering length&  0.0358   &  0.038 $\pm$ 0.003 \\
$a^0_2$ $\pi-\pi$ scattering length&  0.0017   & 0.0017 $\pm$ 0.003\\
$a^2_2$ $\pi-\pi$ scattering length &  -0.0005  & not measured\\
$r_{\pi}$ pion charge radius & 0.55fm &  0.66fm\\ nucleon-core mass  &  1390MeV & 
$\sim$1300MeV  \\ constituent quark rms size   & 0.59fm& -\\ chiral quark
constituent mass &   270MeV&   -\\ u/d quark constituent  mass &  300MeV  & $\sim$
340MeV\\ s quark constituent  mass &  525MeV  & $\sim$510MeV\\
$0^+$ diquark rms size &   0.78fm   & -\\
$0^+$ diquark constituent mass &  692MeV  &  $>$400MeV  \\
$1^+$ diquark constituent  mass &  1022MeV  &  -  \\ 
$0^-$ diquark constituent  mass &  1079MeV  &  -  \\
$1^-$ diquark constituent  mass &  1369MeV  &  -  \\ MIT core bag-constant  & 
(154MeV)$^4$  & (146MeV)$^4$ \\ MIT nucleon-core mass (no cm corr.)  & 1500MeV &  
$\sim$1300MeV   \\
\hline 
\end{tabular}

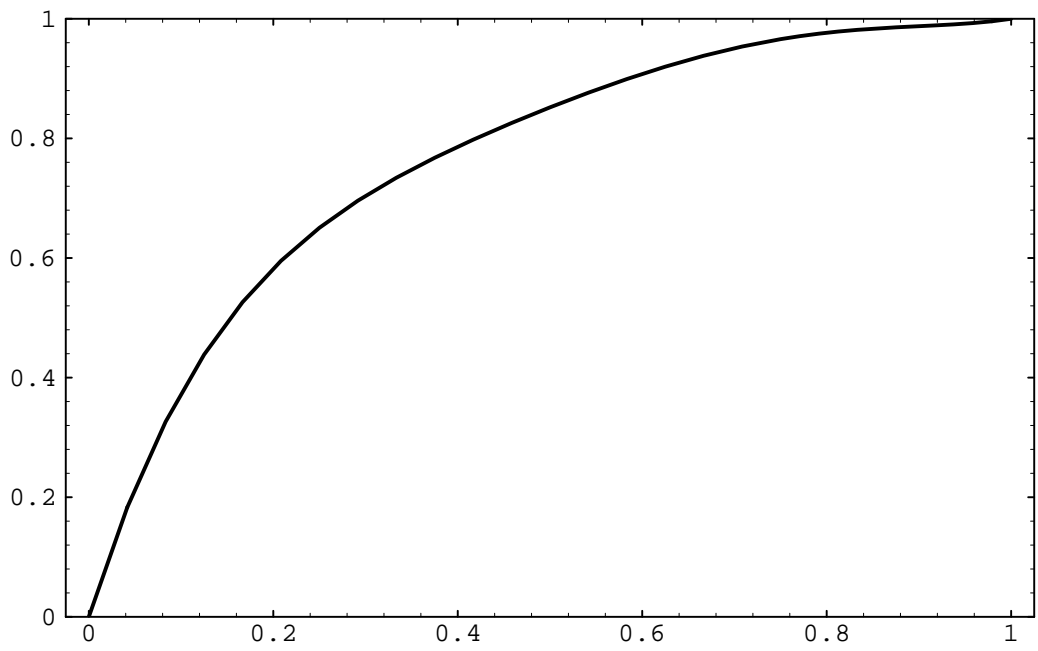
\end{document}